# Dramatic pressure-sensitive ion conduction in conical nanopores

Laetitia Jubin[a,1], Anthony Poggioli[a,1], Alessandro Siria[a], and Lydéric Bocquet[a,2]

[a]Laboratoire de Physique Statistique, Ecole Normale Supérieure, 75005 Paris, France

**Ion transporters in Nature exhibit a wealth of complex transport properties such as voltage gating, activation, and mechanosensitive behavior. When combined, such processes result in advanced ionic machines achieving active ion transport, high selectivity, or signal processing. On the artificial side, there has been much recent progress in the design and study of transport in ionic channels, but mimicking the advanced functionalities of ion transporters remains as yet out of reach. A prerequisite is the development of ionic responses sensitive to external stimuli. In the present work, we report a counterintuitive and highly nonlinear coupling between electric and pressure-driven transport in a conical nanopore, manifesting as a strong pressure dependence of the ionic conductance. This result is at odds with standard linear response theory and is akin to a mechanical transistor functionality. We fully rationalize this behavior on the basis of the coupled electrohydrodynamics in the conical pore by extending the Poisson–Nernst–Planck–Stokes framework. The model is shown to capture the subtle mechanical balance occurring within an extended spatially charged zone in the nanopore. The pronounced sensitivity to mechanical forcing offers leads in tuning ion transport by mechanical stimuli. The results presented here provide a promising avenue for the design of tailored membrane functionalities.**

Ionic transport in nanometric-scale channels and pores has been an intense topic of research for two decades (1, 2), highlighting novel transport processes at the nanoscale (1, 2), with applications to, e.g., physical modeling of biological channels (3, 4), energy generation (1, 5–8), and desalination (9). Coupling of electrostatic and electrodiffusive processes at the nanoscale can result in ionic current rectification, in analogy with classical solid-state semiconductor diodes (1, 10–14). Diode-like behavior of the ionic current can be harvested for applications such as solvent rectification and desalination (9, 15–18), counter-gradient ionic pumping (19), and energy harvesting (20).

However, the ability to tune the ionic conduction by an external stimulus, exhibited in biological nanopores (21–24), remains challenging to achieve artificially. The nanofluidic equivalent of the transistor, pioneered by refs. 25 and 26, still poses many difficulties for efficient implementation in nanofluidic circuitry. More generally, the nonlinear response of ionic transport in nanopores to external forcings and its relation to the nanopore geometry remain poorly understood. Diode-like behavior highlights a rectified current–voltage response in asymmetric nanopores, but the extension to driving forces other than the electric forcing remains largely unexplored up to now. Of particular interest is the coupling of ionic transport to mechanical forcings—e.g., an imposed pressure drop—beyond the linear response regime. The ability to tune the conduction by such an external stimulus would open up the possibility of designing advanced fluidic circuitry, such as the hypothetical memristor response (27). However, the low Reynolds number hydrodynamics governing nanometric fluid transport are by essence time reversible and resolutely linear in pressure forcing, suggesting the a priori impossibility of a nonlinear mechanical response. Accordingly the ionic response to mechanical driving is usually described in terms of a streaming current that is linear in pressure drop and verifies the Onsager symmetry relations (1).

In this paper, we report observations of ionic transport in conical nanocapillaries demonstrating a considerable nonlinear sensitivity of ionic conduction to applied pressure. This results in an ionic conductance that exhibits a nonanalytic increase or decrease with pressure, allowing for robust and versatile tuning of the conductance. Even more striking is the fact that this behavior is observed for small imposed pressure drops, while the classical linear response is recovered at large imposed pressure drops, at odds with the expected linear response scenario. We rationalize these observations on the basis of a theoretical framework based on the coupled electromechanical dynamics in the conical nanopore. It demonstrates the role played by the pressure-induced modifications of the spatially charged zone (SCZ) that forms in the interior of the nanopore and is at the origin of the highly nonlinear current response.

APPLIED PHYSICAL SCIENCES

## Experiments

We conducted experiments on conical glass nanocapillaries, which constitute easily fabricated and readily reproducible physical models of a single conical nanopore. The interior radius of these nanocapillaries varied between 250 μm on the upstream end and $165 \pm 15$ nm at the tip over a length of ∼3 mm, corresponding to a half angle of approximately $5^{\circ}$. (Fig. 1*B* and *SI Materials and Methods*). During the experiments, ionic current was measured as a function of applied voltage ($-400 < \Delta V < +400$ mV) and pressure ($0 < \Delta P < 1,500$ mbar) at a fixed ionic concentration $c_0$ of potassium chloride (KCl), as illustrated in Fig. 1*A* and detailed in *Materials and Methods*.

## Significance

**The domain of nanofluidics, exploring fluid and ionic transport at the nanoscale, has made tremendous progress recently. However, transport in artificial systems remains as yet unable to reproduce the richness and complexity of the advanced functionalities exhibited in Nature, such as mechanosensitive channel activation, ionic pumps, etc. Here we demonstrate experimentally and theoretically a mechanosensitive ionic transistor effect in conical nanopores, in which the ionic conduction is found to be activated or inhibited by a mechanical stimulus. Such behavior is a fundamental building block for reproducing the advanced transport functionalities found in Nature.**

Author contributions: A.S. and L.B. designed research; L.J., A.P., A.S., and L.B. performed research; L.J. and A.P. analyzed data; and L.J., A.P., A.S., and L.B. wrote the paper.

The authors declare no conflict of interest.

This article is a PNAS Direct Submission.

[1]L.J. and A.P. contributed equally to this work.

[2]To whom correspondence should be addressed. Email: lyderic.bocquet@lps.ens.fr.

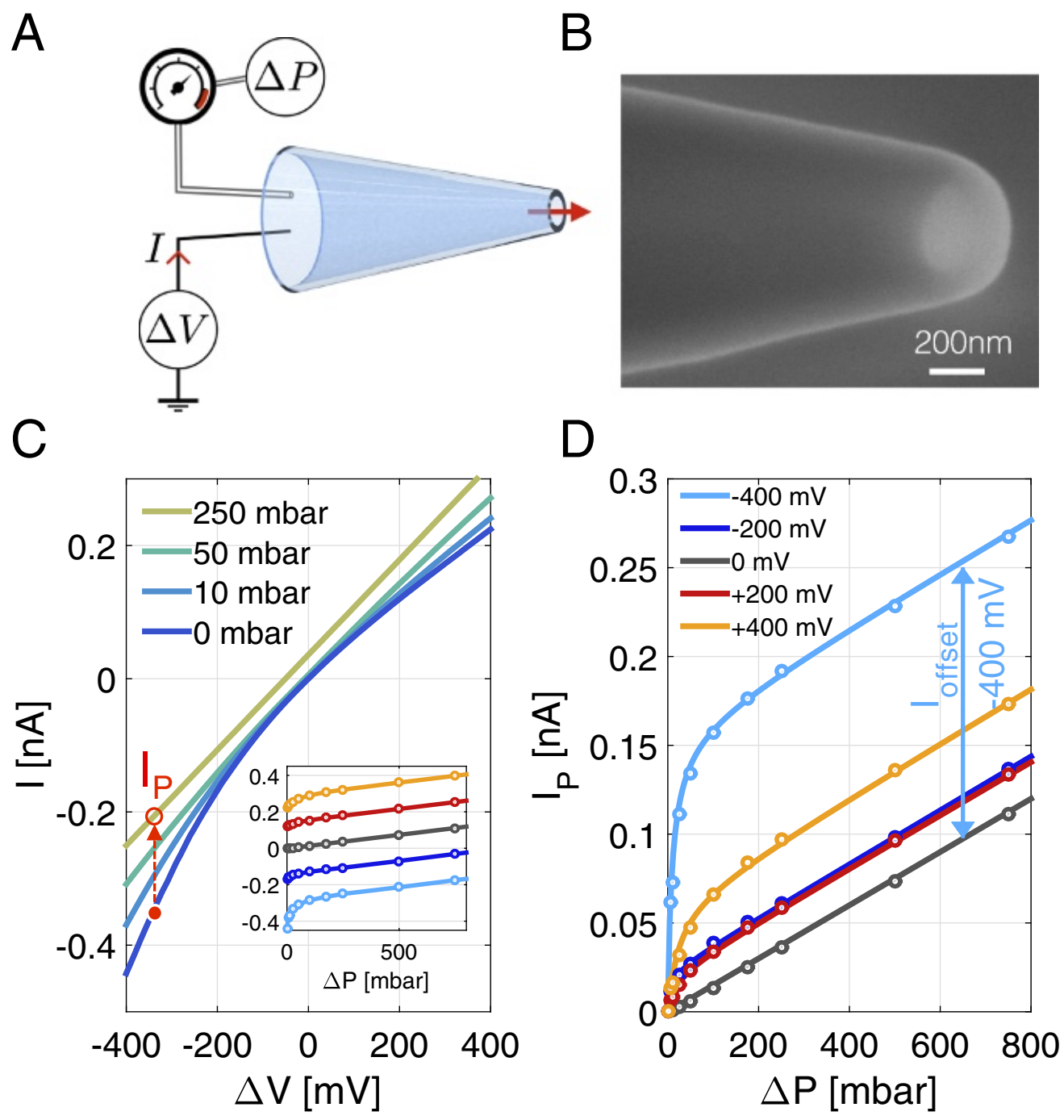

**Fig. 1.** Experimental setup with a single conical nanopore and experimental response of the ionic current $I$ to applied voltage $\Delta V$ and pressure $\Delta P$. (*A*) Sketch. (*B*) SEM image. (*C*) Current–voltage curves for increasing values of $\Delta P$, as indicated in the key. *Inset* shows the current as a function of $\Delta P$ for several different values of $\Delta V$, colored according to the key in *D*. (*D*) Additional current induced by applied pressure, $I_P$, as a function of $\Delta P$ for several different values of $\Delta V$, as indicated in the key. The arrow indicates $I_{\text{offset}}$, the offset in $I_P$ compared with the linear response obtained for $\Delta V = 0$, for $-400$ mV. The experimental data are fitted according to Eq. **1** (solid lines). All measurements correspond to a molarity $[\text{KCl}] = 10^{-3}$ M, pH $\simeq 6$, and a nominal tip radius of $R_0 = 165 \pm 15$ nm.

The results of experiments conducted at a molarity $[\text{KCl}] = 10^{-3}$ M and pH $\approx 6$ are shown in Fig. 1 *C* and *D*. At this concentration and pH, we observe substantial rectification of the current–voltage (IV) curve for $\Delta P = 0$ (Fig. 1*C*), in agreement with previous observations (17, 28, 29). Conversely, for $\Delta V = 0$, the pressure-driven response behaves as expected and a streaming current is generated, linear in $\Delta P$ (Fig. 1*C*, *Inset*) (1, 2, 30). This streaming current originates from the pressure-induced advection of ions within the Debye screening layer that forms in the vicinity of solid–liquid interfaces (2). It is given by the Smoluchowski result, which, in a conical nanopore with linearly varying radius of slope $\alpha_1$, takes the form $I_{\text{stm}} = \pi R_0 \alpha_1 \mu_{\text{EO}} \times \Delta P$, where $R_0$ is the minimum radius occurring at the tip of the nanopore, and $\mu_{\text{EO}} \equiv (\epsilon/\eta)(-\zeta)$ is the electroosmotic mobility. In the latter quantity, $\zeta$ is the so-called zeta potential, and $\eta$ and $\epsilon$ are the water viscosity and dielectric constant, respectively. Quantitatively, the streaming conductance $\text{S}_{\text{stm}} = I_{\text{stm}}/\Delta P$ obtained at zero applied voltage is $\text{S}_{\text{stm}}^{\text{exp}} = 0.153\,\text{nA} \cdot \text{bar}^{-1}$, corresponding to a zeta potential of $\zeta \simeq -42$ mV in agreement with the literature (31). Incidentally, by comparing the above expression for the streaming current to the classical result obtained for a cylindrical geometry, $I_{\text{stm}} = (\pi R_0^2/L)\mu_{\text{EO}} \times \Delta P$, we find that the gradient is confined to an effective length of the order $L_{\text{eff}} = R_0/\alpha_1$.

The behavior for combined finite $\Delta V$ and $\Delta P$ differs dramatically from the behavior observed when only one forcing is nonzero. As shown in Fig. 1*C*, the IV response changes qualitatively for increasing applied pressures, linearizing for pressures $\gtrsim 200$ mbar. Similarly, for a given applied voltage drop $\Delta V$, the current–pressure (IP) response is dramatically nonlinear for small pressures $\Delta P \lesssim 50$ mbar. As shown in Fig. 1*D*, this nonlinearity is particularly apparent if we examine the additional current induced by pressure, $I_P \equiv I(\Delta P, \Delta V) - I(\Delta P = 0, \Delta V)$, as a function of $\Delta P$ for fixed values of $\Delta V$. As shown in Fig. 1*D*, for any voltage drop the pressure dependence of $I_P$ is well described by a simple expression of the form

$$I_P(\Delta P) = S_{\text{stm}} \Delta P + I_{\text{offset}} \frac{a_1\, \Delta P^{1/2} + a_2\, \Delta P}{1 + a_1\, \Delta P^{1/2} + a_2\, \Delta P}, \qquad [1]$$

where the fitting coefficients $a_i$ and $I_{\text{offset}}$ are functions of the voltage drop $\Delta V$. This highlights a small pressure response of the current that scales as $I_P \sim \Delta P^{1/2}$, while the linear regime $I_P \sim \Delta P$ is recovered for large pressure. The square-root dependence of the ionic current on pressure as $\Delta P \to 0$ suggests that the response is nonanalytic at $\Delta P = 0$, within the accuracy of the experiments. This response is obviously at odds with naive considerations, which would rather suggest that the pressure-induced response at small $\Delta P$ should take the form of a Taylor expansion in odd powers of $\Delta P$, $I_P \simeq b_1 \Delta P + b_3 \Delta P^3 + \ldots$, where the coefficients of the expansion may themselves be expressed as (even) analytic expansions in $\Delta V$. As highlighted by Fig. 1*D*, the response to mechanoelectric driving forces is highly nonlinear and far stronger than such considerations would suggest.

Surprisingly, while the pressure response observed here is highly nonlinear for small $\Delta P$ (and any finite $\Delta V$), the limiting slope of the IP curves obtained for large $\Delta P$ is independent of voltage and equal to the slope obtained when $\Delta V = 0$ V (Fig. 1*D*). In this large $\Delta P$ regime, the IP response is again characterized by a linear relationship, but now with a voltage-dependent offset current (Eq. **1**). The offset current represents a substantial enhancement of the streaming current that would be obtained in the ordinary linear response regime; this can be seen by comparing the linear IP curves corresponding to $\Delta V = 0$ and $-400$ mV (Fig. 1*D*).

Finally, we note that this drastically nonlinear behavior is obtained when no Debye-layer overlap occurs in the nanocapillary; the Debye layer here is $\sim$10 nm thick, an order of magnitude smaller than the minimum nanocapillary radius, $R_0 = 165 \pm 15$ nm.

Altogether, these results are best described in terms of a pressure-dependent ionic conductance. We report in Fig. 2 *A* and *B*, respectively, both the apparent conductance $G_{\text{app}} \equiv \Delta I/\Delta V$ and the differential conductance $G_{\text{diff}} \equiv \partial I/\partial \Delta V$. Both quantities highlight a strong sensitivity of the conductance to pressure for small applied pressures. Depending on the applied voltage, the conductance increases or decreases with $\Delta P$, with variations of up to 100% for a change of pressure as small as 100 mbar. The result is an ionic conduction that is dramatically dependent on external mechanical conditions.

## Model

To rationalize the a priori unexpected experimental results presented above, we have developed a one-dimensional model of the ionic transport based on the radially integrated Poisson, Nernst–Planck, and Stokes equations for the electrostatic field, ionic fluxes and concentrations, and the pressure and hydrodynamic velocity, respectively. Radial integration of the Nernst–Planck and Stokes equations typically proceeds on the assumption of a local Poisson–Boltzmann equilibrium (1, 2, 7). Such an equilibrium requires local electroneutrality, with the integrated ionic charge density exactly balancing the local surface charge density everywhere. However, in the present model, we allow for small deviations from local equilibrium with the possibility of nonelectroneutral regions, so-called SCZs. These deviations are characterized by a local net charge $\delta n_c$, defined as

$$\delta n_c \equiv n_c + \mathrm{S}\frac{2\mathrm{Du}_0}{R}, \qquad [2]$$

where $n_c$ is the local volume density of ionic charge, nondimensionalized by $ec_0$; $e$ is the elementary charge; $c_0$ is the ionic concentration in the reservoirs; $\mathrm{S}=-1$ is the sign of the surface charge for glass (31); $R$ is the local radius, nondimensionalized by the minimum value $R_0$; and the Dukhin number is defined in terms of the magnitude of the surface charge $|\sigma|$, the reservoir concentration, and the minimum radius as

$$\mathrm{Du}_0 \equiv \frac{|\sigma|}{ec_0 R_0}. \qquad [3]$$

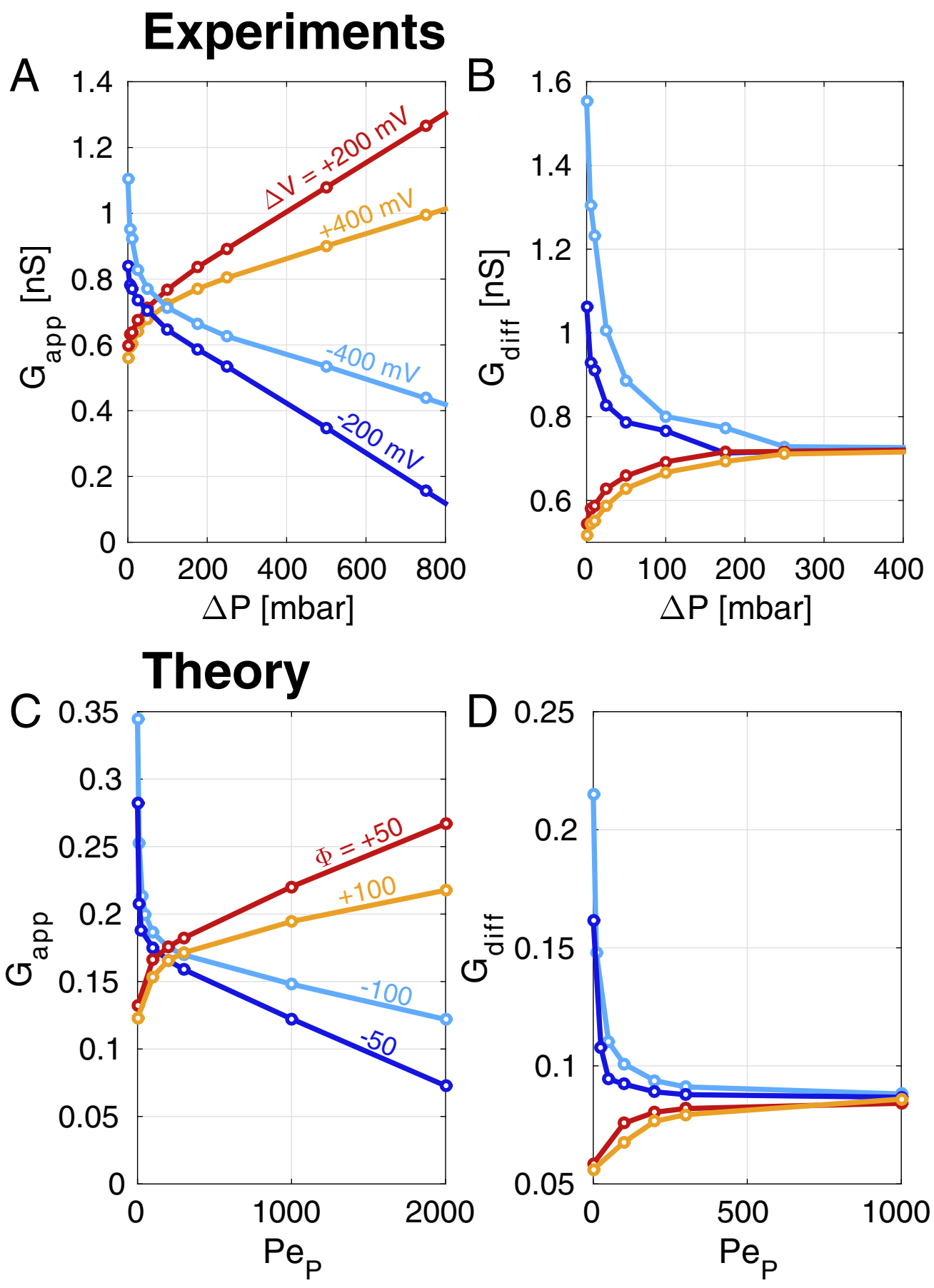

**Fig. 2.** Experimental (*A* and *B*) and model-derived (*C* and *D*) effective (*A* and *C*) and differential (*B* and *D*) conductance as a function of applied pressure for several different values of applied voltage. The corresponding values of $\Delta V$ ($\Phi$) are indicated in *A* (*C*). $\Phi$ and $Pe_P$ are the dimensionless rescaled voltage and pressure, respectively, and are defined in the main text.

The Dukhin number indicates the relative importance of ionic transport in the screening layer vs. in the bulk (1). In everything that follows, we use the nondimensionalized variables defined in Table 1.

The dimensionless and radially averaged 1D fluxes take the form

$$I = \overbrace{\pi R^2\left(-\frac{dn_c}{dx}\right)}^{I_{\mathrm{diff}}} + \overbrace{\pi R^2 c\left(-\frac{d\phi}{dx}\right)}^{I_{\mathrm{ep}}} + Q\delta n_c + \overbrace{2\pi Rk\left(-\frac{d\phi}{dx}\right)}^{I_{\mathrm{eo}}} + \overbrace{\frac{1}{\mathrm{Pe}_{\mathrm{osm}}}\frac{\mu_{\mathrm{EO}}}{\mu_{\mathrm{EP}}}\pi R^2\left(-\frac{dP}{dx}\right)}^{I_{\mathrm{stm}}} + \overbrace{2\pi Rk\left(\mathrm{S}\frac{d\ln c}{dx}\right)}^{I_{\mathrm{do}}} \qquad [4]$$

$$J_{\mathrm{sol}} = -\pi R^2\left(\frac{dc}{dx} + n_c\frac{d\phi}{dx}\right) + Qc, \qquad [5]$$

$$Q = -\pi R^4\left(\frac{dP}{dx} + \mathrm{Pe}_{\mathrm{osm}}\delta n_c\frac{d\phi}{dx}\right) + \pi R^2\left[\frac{\mu_{\mathrm{EO}}}{\mu_{\mathrm{EP}}}\left(-\frac{d\phi}{dx}\right) + \frac{\mu_{\mathrm{DO}}}{D}\left(-\frac{d\ln c}{dx}\right)\right], \qquad [6]$$

which verify conservation laws

$$\frac{dI}{dx} = \frac{dJ_{\mathrm{sol}}}{dx} = \frac{dQ}{dx} = 0, \qquad [7]$$

together with the Poisson equation

$$\left(\frac{\lambda_D}{\ell}\right)^2 \frac{1}{\pi R^2}\frac{d}{dx}\left(\pi R^2\frac{d\phi}{dx}\right) + \delta n_c = 0. \qquad [8]$$

These equations are derived in *Derivation of Eqs. 4–8: Radial Integration of PNPS Equations*; they express conservation of electric charge, solute mass, and solvent volume. In the right-hand side of Eq. **4**, the terms represent the contributions from diffusion ($I_{\mathrm{diff}}$), electrophoresis ($I_{\mathrm{ep}}$), bulk advective transport of the local net charge, electroosmosis ($I_{\mathrm{eo}}$), streaming current ($I_{\mathrm{stm}}$), and diffusioosmosis ($I_{\mathrm{do}}$). Similar interpretations hold for the terms in Eq. **5**. We emphasize that we do not presuppose the overlap of Debye layers.

The mechanisms at play are described in terms of various mobilities: $\mu_{\mathrm{EP}}$ and $\mu_{\mathrm{EO}}$, the electrophoretic and -osmotic mobilities; $\mu_{\mathrm{DO}}$, their diffusioosmotic counterpart (32); and the diffusion coefficient $D$, related to the electrophoretic mobility by the Einstein relation $D=\mu_{\mathrm{EP}}\times k_B T/e$. The equations also contain the Debye length $\lambda_D \equiv \sqrt{\epsilon_r\epsilon_0 k_B T/e^2 c_0}$ and $\ell$, the geometric length scale characterizing the nanocapillary tip (Fig. 3). In addition to the Dukhin number (Eq. **3**), we have introduced the dimensionless parameter

$$\mathrm{Pe}_{\mathrm{osm}} \equiv \frac{R_0^2 k_B T c_0}{8\eta D}, \qquad [9]$$

which compares diffusion to a pressure-driven flow associated with the scale of an osmotic pressure $k_B T c_0$ and may accordingly be interpreted as an osmotic Péclet number. Finally, $2\pi Rk$

**Table 1. Model variables and their rescaled dimensionless counterparts**

| Variable | Notation | Rescaled |
|---|---|---|
| Position | $x$ | $x \to \ell x$ |
| Radius | $R$ | $R \to R_0 R$ |
| Concentration | $c$ | $c \to c_0 c$ |
| Charge Density | $n_c$ | $n_c \to e c_0 n_c$ |
| Potential | $\phi$ | $\phi \to (k_B T/e)\phi$ |
| Solute flux | $J_{\mathrm{sol}}$ | $J_{\mathrm{sol}} \to (R_0^2 D c_0/\ell) J_{\mathrm{sol}}$ |
| Electric Current | $I$ | $I \to (R_0^2 e D c_0/\ell) I$ |
| Water flux | $Q$ | $Q \to (R_0^2 D/\ell) Q$ |
| Pressure | $P$ | $P \to (8\eta D/R_0^2) P$ |

The length $\ell$ characterizes the transition between the interior of the nanopore and the reservoir (Fig. 3 and main text).

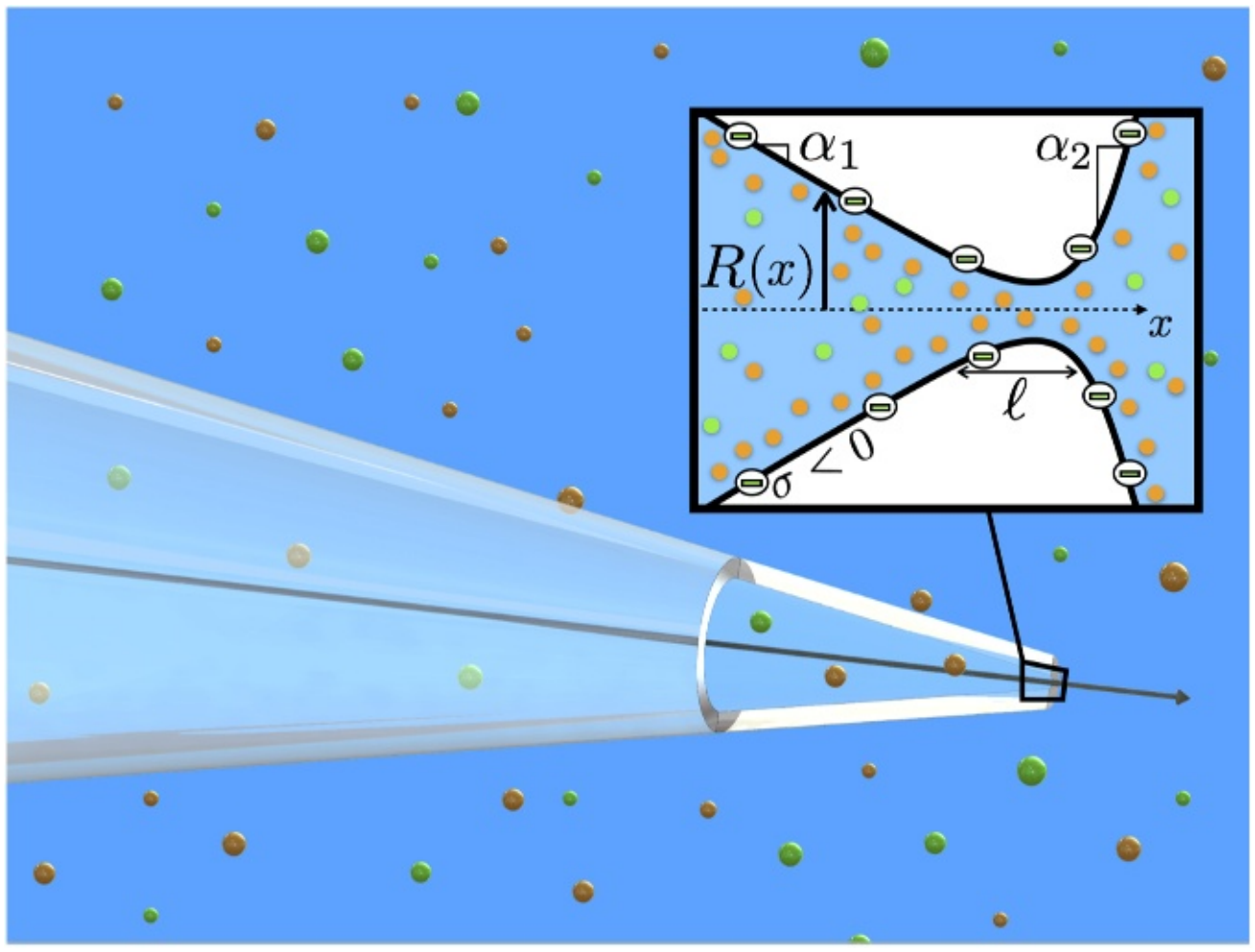

**Fig. 3.** A sketch of the geometry of the glass nanocapillary. *Inset* shows a zoom-in of the model geometry in the vicinity of the tip. The radius is taken to vary between two regions of linear variation over a length-scale $\ell$ as indicated, and these regions are characterized by radial slopes $\alpha_1 = 0.1$ in the interior and $\alpha_2 = 200$ in the exterior.

is the local cross-sectionally integrated electroosmotic conductivity, with $k \equiv 16\mathrm{Du}_0\mathrm{Pe}_{\mathrm{osm}}(\lambda_D/R_0)^2$. The values of all of the dimensionless parameters governing the solution were estimated for the experiments presented above: $\mathrm{Du}_0 \simeq 0.5$, $\lambda_D/R_0 \simeq 0.05$, $\lambda_D/\ell \simeq 0.1$, $\mu_{\mathrm{EO}}/\mu_{\mathrm{EP}} \simeq 0.2$, $\mu_{\mathrm{DO}}/D \simeq 1$, $\mathrm{Pe}_{\mathrm{osm}} \simeq 10$, and $\ell/R_0 \approx 0.5$. Additionally, the interior radial slope is $\alpha_1 \approx \tan(5^\circ) \approx 0.1$. These parameters are used in all of the numerical calculations presented below.

The model geometry is shown in Fig. 3. The radius is taken to vary continuously between two linear regions: The upstream region of slope $dR/dx = -\alpha_1$, representing the nanopore interior, and the downstream region of slope $dR/dx = +\alpha_2 = 200$, representing the rapid divergence of the radius as the downstream reservoir is approached. The transition between radial slopes occurs continuously over a length $\ell$ (Fig. 3 and *Model Geometry*). The boundary conditions are imposed in the reservoirs at $x = \pm\infty$: $\phi(x \to +\infty) = P(x \to +\infty) = 0$, $\phi(x \to -\infty) = \Phi$, $P(x \to -\infty) = \mathrm{Pe}_P$, together with conditions on the charge and salt concentration $c(x \to \pm\infty) = 1$ and $n_c(x \to \pm\infty) = 0$. In the preceding, we have introduced the dimensionless driving forces

$$\Phi \equiv e\Delta V/k_B T \approx \Delta V/25\,\mathrm{mV}$$
$$\mathrm{Pe}_P \equiv R_0^2 \Delta P/8\eta D \approx \Delta P/6\,\mathrm{mbar}. \qquad [10]$$

## Results

The coupled transport equations given above, Eqs. **4–7**, were solved numerically, and the results for the response of the current to applied voltage and pressure are reported in Fig. 4. The results for the conductance are reported in Fig. 2 *C* and *D*.

Crucially, this theoretical framework reproduces all of the essential qualitative features of the experiments. Comparing the experimental results—Figs. 1 *C* and *D* and 2 *A* and *B*—to the theoretical predictions—Figs. 2 *C* and *D* and 4—, we see that we successfully recover a strong, nonlinear dependence of the ionic conduction on pressure (Fig. 2 *C* and *D*), resulting in a highly sensitive response of the pressure-induced current $I_P$ to pressure (Fig. 4*B*). We note that, in the model, the applied forcings are larger than those in the experiments. This was necessary to recover the correct degree of rectification in the IV curves and is presumably due to the simplifying one-dimensionality of our model.

Furthermore, the prediction for the Péclet dependence of the current $I_P$ shown in Fig. 4*B* is successfully described by Eq. **1**, in full agreement with the experimental results shown in Fig. 1*D*. This demonstrates that the experimental behavior $I_P \sim \Delta P^{1/2}$ measured for low pressure drop is fully recovered by the model, indicating a strong sensitivity to applied pressure for small pressures. Finally, the theoretical IV curves are observed to linearize as pressure is increased, in accordance with the experimental observations shown in Fig. 1*C*, with the conductance for all voltages approaching the conductance at zero voltage drop, $\Phi = 0$ (Fig. 4*A*). The apparent offset in the linear streaming current for large applied pressures is asymmetric in applied voltage, growing much more quickly for negative than for positive values of $\Phi$, in full agreement with its experimental counterpart.

## Discussion: The Deformation of the SCZ

We now show that this nontrivial behavior originates in the sensitivity of the SCZ to the balance between electrical and mechanical forcing. We first note that at equilibrium the present system exhibits a nonvanishing net charge density ($\delta n_c$, Eq. **2**) profile. Solving the above transport equations for $I = J_{\mathrm{sol}} = Q = 0$, we find that the net charge density at equilibrium, $\delta n_c^{\mathrm{equ}}$, is nonzero in the conical system and obeys an implicit linear relationship with the Dukhin number,

$$G_b\left(\frac{\ell}{\lambda_D}\right)^2 \int_{-\infty}^{+\infty} \frac{\mathrm{d}x}{\pi R^2} \int_{-\infty}^{x} \mathrm{d}x' \pi R^2 \delta n_c^{\mathrm{equ}} = 2\pi \frac{\ell}{R_0} \alpha_1 \mathrm{S}\,\mathrm{Du}_0, \qquad [11]$$

where $G_b = \left[\int_{-\infty}^{+\infty} \mathrm{d}x/\pi R^2\right]^{-1}$ is the (dimensionless) bulk electrophoretic conductance. (Note that a more detailed derivation of this equation and those following is given in *Derivation of Eqs. 11 and 12*, *Derivation of Eq. 13*, and *Derivation of Eq. 14*.) This result may be interpreted in terms of the buildup of a Donnan potential inside the conical nanocapillary, which in the present conditions (a nonoverlapping Debye layer) disappears as the capillary angle $\alpha_1$ vanishes and a simple cylindrical geometry is obtained.

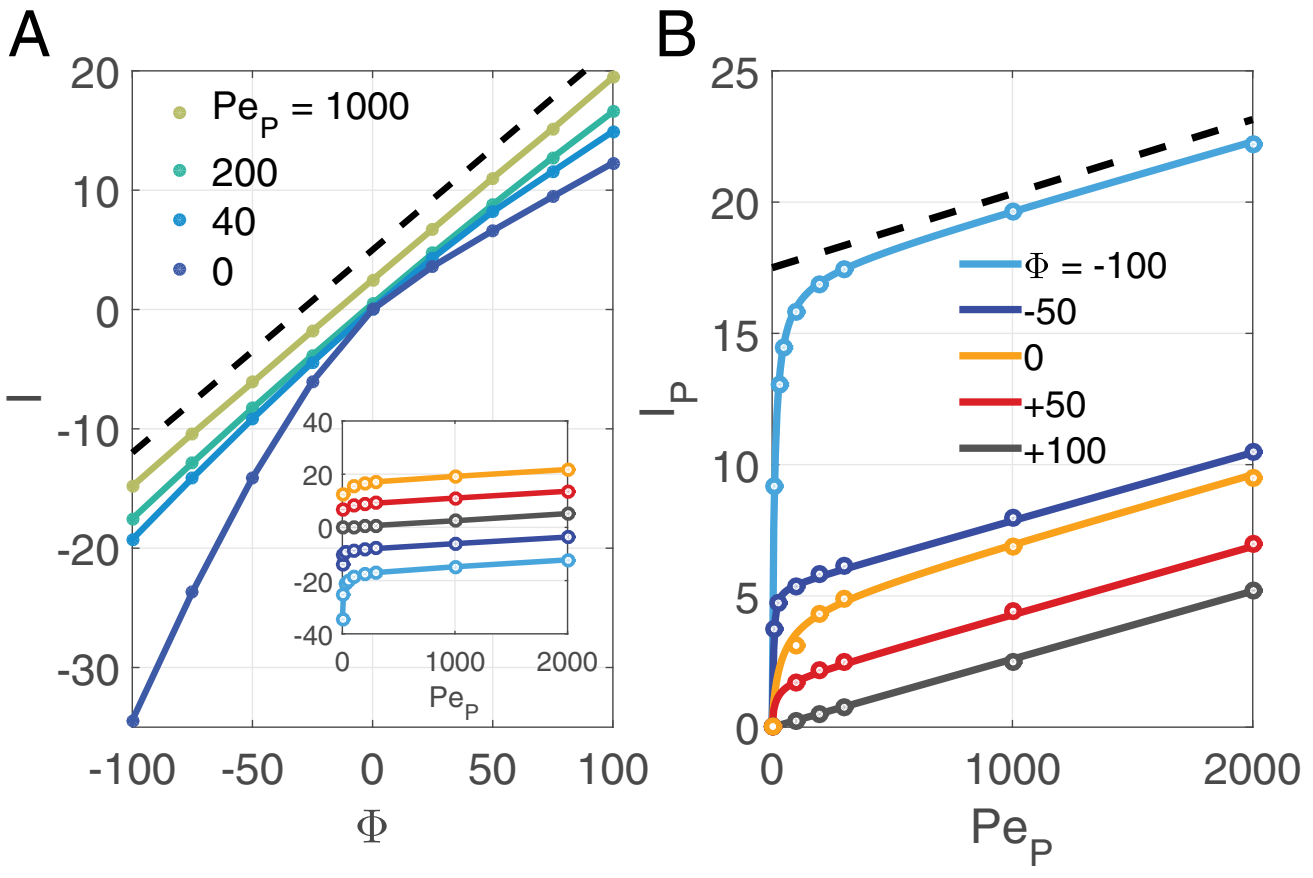

**Fig. 4.** Model-derived response of the ionic current $I$ to applied voltage $\Phi$ and pressure $\mathrm{Pe}_P$ in a conical quasi–one-dimensional geometry (Fig. 3). (*A*) IV curves for increasing values of $\mathrm{Pe}_P$, as indicated in the key. *Inset* shows the current as a function of $\mathrm{Pe}_P$ for several different values of $\Phi$, colored according to the key in *B*. (*B*) Additional current induced by applied pressure, $I_P$, as a function of $\mathrm{Pe}_P$ for several different values of $\Phi$, as indicated in the key. In *B*, the model predictions are fitted according to Eq. **1**, similar to Fig. 1 for the experimental data (with $\Delta P$ replaced by the Péclet number $\mathrm{Pe}_P$). The slope of the dashed black line in *A* indicates the value of $G_0$, and that in *B* indicates the value of $S_{\mathrm{stm}}$ (Eq. **14**).

Under applied voltage and pressure drops, this equilibrium SCZ is modified. Interestingly, one may still obtain an explicit integral relationship between the current and the charge imbalance:

$$I = G_b \left[ \Phi - \left(\frac{\ell}{\lambda_D}\right)^2 \int_{-\infty}^{+\infty} \frac{\mathrm{d}x}{\pi R^2} \int_{-\infty}^{x} \mathrm{d}x' \pi R^2 \left[\delta n_c - \delta n_c^{\mathrm{equ}}\right]\right]. \quad [12]$$

In practice, this equation is obtained by first solving the Poisson equation (Eq. **8**) for the electric flux at $x=-\infty$ and then relating this to the ionic current by examining the limiting behaviors of the flux equations (Eqs. **4–6**).

This result explicitly confirms that the nonlinear response results from the deformation of the SCZ under these driving forces. From the model results, we learn that for small Péclet number $\mathrm{Pe}_P \to 0$, the current is dominated locally by the electrophoretic and electroosmotic responses. On the other hand, at large Péclet number, the linearization of the IP response is found to correspond to an increase in importance of the local streaming current, such that the current is dominated by the local electrophoretic, electroosmotic, and streaming current responses. We discuss each of these two regimes in turn.

**Small Péclet Regime.** In this regime, the striking result is the dramatic sensitivity of the conductance to applied pressure. As suggested by Eq. **1**, the current $I_P$ exhibits a nonanalytic square-root dependence on pressure drop (or Péclet number) in this regime, so that the model predicts $G \sim \Delta P^{1/2}$. The conductance in this small Péclet regime may be estimated by retaining only the electrophoretic and electroosmotic terms in the expression for the ionic current (Eq. **4**). Integrating in $x$, one finds for the apparent conductance $G = I/\Delta V$,

$$G = \left[\int_{-\infty}^{+\infty} \frac{\mathrm{d}x}{\pi R^2 \left(c + \frac{2k}{R}\right)}\right]^{-1}, \quad [13]$$

where the (nonlinear) pressure dependence is hidden in the concentration profile $c(x;\mathrm{Pe}_P,\Phi)$. The above result for the apparent conductance (Eq. **13**) shows that for small Péclet number, the variation in the apparent conductance with pressure drop, $G(\mathrm{Pe}_P)$, may be understood in terms of the pressure-induced variations in the concentration profile $c(\mathrm{Pe}_P)$ relative to equilibrium. Concentration profiles are plotted in *Concentration and Cumulative Charge Profiles* and Fig. S3. These profiles exhibit a strong sensitivity to applied pressure when $\mathrm{Pe}_P < 50$, with the concentration everywhere relaxing to the reservoir value as the linearizing Péclet number $\mathrm{Pe}_P^{\mathrm{lin}} \approx 300$ is approached.

The equivalence of Eq. **13** and the more general Eq. **12** in describing the current response at low $\mathrm{Pe}_P$ implies a direct relationship between the net spatial charge $\delta n_c$ and the concentration field. This is illustrated in Fig. S3, where we have also plotted profiles of the cumulative charge in the nanopore, $\delta q \equiv \int_{-\infty}^{x} \mathrm{d}x \pi R^2 \delta n_c$. The structural similarity between the cumulative charge and the excess concentration relative to the reservoir value $\delta c \equiv c - 1$ is immediately apparent.

This suggests a back-of-the-envelope argument to account for the square-root variation of the conductance with Péclet number, $G \sim \mathrm{Pe}_P^{1/2}$, highlighted in Eq. **1**. Under a pressure drop, one may anticipate a simple mechanical balance for the SCZ between the electrostatic and pressure forces. This typically takes the form $\delta q \times E_{\mathrm{app}} \sim R_0^2 \Delta P$, where $\delta q$ is the variation of the cumulative charge in the SCZ and $E_{\mathrm{app}}$ is the variation in the induced electric field, under the applied pressure. Equivalently, one may interpret the force balance in terms of a balance between the Maxwell stress and the applied pressure: $\frac{1}{2}\epsilon E_{\mathrm{app}}^2 \sim \Delta P$. The equivalence of these perspectives requires that the cumulative charge in the SCZ be proportional to the induced electric field, $\delta q \sim E_{\mathrm{app}}$, in agreement with the Poisson equation, Eq. **8**. Solving for $\delta q$ yields $\delta q \sim \pm\sqrt{\Delta P}$ (depending on the sign of the applied voltage drop) or, in dimensionless variables, $\delta q \sim \pm\sqrt{\mathrm{Pe}_P}$. Furthermore, as noted above, the variation in the concentration field is found to scale with the cumulative charge. We thus have $\delta c \sim \delta q \sim \pm\sqrt{\mathrm{Pe}_P}$. From Eq. **13**, this variation in the concentration leads to a modification of the conductance scaling as $\delta G(\mathrm{Pe}_P) \sim \pm\sqrt{\mathrm{Pe}_P}$ for small Péclet number.

While this scaling is established a posteriori, it recovers the nonanalytic square-root correction to the conductance in the Péclet number, in full agreement with the corresponding variation observed in the experiments and theory, as shown in Fig. 2. It also suggests that the origin of the nonanalyticity is in the contribution to the mechanical balance on the SCZ of the Maxwell stress tensor and its quadratic dependence on the electric field. Interestingly, this dependence is expressed for the present conical geometry, but disappears for cylindrical geometries with constant radius.

**Large Péclet Regime.** In this regime, the current response is linear in applied voltage and pressure (Figs. 1 and 4). This puzzling behavior can be rationalized analytically by noting that the current is dominated by the electrophoretic, electroosmotic, and streaming contributions. Under these conditions, one can deduce the analytical expression for the current as

$$I = G_0 \left(\Phi + \gamma \frac{\mu_{\mathrm{EO}}}{\mu_{\mathrm{EP}}} \frac{\mathrm{Pe}_P}{\mathrm{Pe}_{\mathrm{osm}}}\right) \equiv G_0 \Phi + \mathrm{S}_{\mathrm{stm}} \mathrm{Pe}_P, \quad [14]$$

where $\gamma \equiv \left[\int_{-\infty}^{+\infty} \mathrm{d}x/\pi R^4 \left(1+\frac{2k}{R}\right)\right] \Big/ \left(\int_{-\infty}^{+\infty} \mathrm{d}x/\pi R^4\right)$ is an order-one constant. We have also introduced $G_0 = \left[\int_{-\infty}^{+\infty} \mathrm{d}x/\pi R^2 \left(1+\frac{2k}{R}\right)\right]^{-1}$, the bulk electrophoretic conductance $G_b$ enhanced by the electroosmotic contribution (1).

This result, although formally clear, is striking in several aspects. First, in this regime, the current is shown to be linear in both voltage drop ($\Phi$) and pressure drop ($\mathrm{Pe}_P$), in full agreement with the experimental results. Furthermore, the conductance and streaming current take their linear response values, i.e., those calculated at vanishing voltage and pressure drop. Indeed, with the exception of small corrections induced by the inclusion of the electroosmotic contribution, the streaming conductance $\mathrm{S}_{\mathrm{stm}} \equiv \partial I/\partial \mathrm{Pe}_P = G_0 \gamma (\mu_{\mathrm{EO}}/\mu_{\mathrm{EP}})(1/\mathrm{Pe}_{\mathrm{osm}}) \approx G_b (\mu_{\mathrm{EO}}/\mu_{\mathrm{EP}})(1/\mathrm{Pe}_{\mathrm{osm}})$ takes the value obtained by assuming that the streaming current alone drives the local current response at high pressure everywhere in the nanopore.

## Conclusions and Perspectives

Our results demonstrate that the ionic conductance of a conical nanopore can be tuned very sensitively by the applied pressure. An increase or decrease of the conductance by up to 100% can be achieved under slight variations in the pressure, stimulating a transition from a high to a low conductance state (or vice versa). Furthermore, the pressure dependence of the conductance is found to be nonanalytical in the applied pressure drop, while a linear behavior is counterintuitively recovered only at high pressure. This strongly nonlinear transport behavior is fully captured by a theoretical framework accounting for the deformation of the SCZs under the coupled mechanical and electrical drivings.

The possibility to mechanically tune conduction mimics a mechanical transistor functionality, with the pressure opening or closing conductance channels. In organisms, the mechanosensitive response of dedicated pores is of crucial importance in preventing fatal osmotic shocks by activating channels under hydrostatic or diffusioosmotic forces (21–23). Here the response to pressure is of a different type, as the pressure is found to tune rather than open or close the electric conduction channels. However, it would be interesting to extend the present study to the osmotic response induced by a difference in salinity across the nanopore. Such a salt concentration difference results in a Nernst potential difference which, according to our present analysis, will couple to the pressure response. Furthermore, salinity differences lead to diffusioosmotic forces which will enter the mechanical balance on top of electrical and mechanical forces studied here (23). This suggests that more complex nonlinear responses are expected in such situations, opening the possibility to tune both ionic and solute flux responses under a variety of stimuli.

We conclude by noting that such a mechanosensitive behavior may also find applications in the context of membrane science, where the possibility to activate or inhibit the electric conduction by small pressure stimuli could be of interest for various applications. For example, in the context of osmotic (blue) energy, the maximum achievable power is proportional to the electric resistance of the membrane (20), and the present nonlinear couplings may allow for mechanical tunability of the extracted power.

Altogether the pressure-sensitive conduction constitutes an elementary building block which we hope will allow for development of new active functionalities mimicking the advanced machines (33) existing in Nature.

## Materials and Methods

Nanocapillary profiles and tip diameters were determined using scanning electron microscopy (SEM). (Fig. 1*B* and Fig. S1). During the experiments, both ends of the nanocapillary were submerged in reservoirs containing Ag/AgCl electrodes and the same aqueous solution of KCl. Ionic currents were then measured between the electrodes for fixed applied voltages (applied via the upstream electrode; Fig. 1*A* and Fig. S2) and pressures (applied in the upstream reservoir). The current response was recorded in the range $-400 < \Delta V < +400$ mV and $0 < \Delta P < 1,500$ mbar. Further details of the experimental procedure and setup are given in *SI Materials and Methods*.

**ACKNOWLEDGMENTS.** All authors acknowledge funding from the Agence Nationale de la Recherche project BlueEnergy. A.S. acknowledges funding from the European Union's Horizon 2020 Framework Program/European Research Council Starting Grant 637748–NanoSOFT. A.P. acknowledges funding from the European Union's Horizon 2020 Framework Program/European Training Program 674979–NanoTRANS.

# Supporting Information

## Jubin et al. 10.1073/pnas.1721987115

### SI Materials and Methods

In this paper we present the ionic current response of a semi-infinite conical nanochannel under applied voltage and pressure. In this section we describe our experimental setup and methodology.

The conical nanopipette is obtained by locally heating and simultaneously pulling a 10-cm–long borosilicate (BSi) capillary of 0.5 mm inner and 1 mm outer diameter with a Sutter Instruments P-2000 pipette puller. The cylindrical capillary is separated into two nanopipettes, each having an inner diameter of 0.5 mm on one end and on the order of hundreds of nanometers on the other end, depending on the pulling parameters. The data presented in the main text were obtained with a nanopipette of half angle of $\sim 5^\circ$ and a 330-nm inside diameter at the tip, as shown on the SEM images in Fig. S1, pulled over two cycles with the following parameter settings for the pipette puller: heat, 350-300; velocity, 30-20; filament, 2-2; delay, 128-128; pull, 0-100.

Before beginning the experiments, the nanopipette is filled with an ionic solution of KCl in Milli-Q (Millipore) water at a concentration of $10^{-3}$ M and natural pH $\approx 6$. Both ends of the nanopipette are then installed into reservoirs filled with the same KCl solution. The nanopipette is the bridge between the two reservoirs, as illustrated schematically in Fig. S2.

To perform low-noise current acquisition under electrical forcing, a Ag/AgCl electrode is inserted in each reservoir and connected through a cooled headstage amplifier to an Axopatch 200B controlled via LabView and used to impose a voltage on the working electrode in the pipette. The other electrode is connected to ground. This setup allows us to measure the ionic current with 0.1 pA precision.

During the experiments, a pressure between 0 mbar and 1,500 mbar is imposed by an Elveflow pressure generator in the reservoir with the working electrode. The reservoir is connected to the millimetric end of the nanopipette which is referred to as the upstream reservoir, corresponding to the direction of water flow in the presence of a nonzero applied pressure. To measure the response of the ionic current to an applied voltage, we perform triangular voltage cycles with 0-V mean, 400-mV amplitude, 10-mV steps, and a 1-s waiting time. The system is observed to rapidly reach a steady state, with a hysteresis in the upward and downward voltage ramps of less than 10% of the mean current value. IV curves are measured for different pressure drops. The hysteresis is observed to disappear reversibly as pressure is increased. The ionic current measurements are obtained by averaging the upward and downward ramps.

### Concentration and Cumulative Charge Profiles

The concentration $c$ and cumulative charge $\delta q \equiv \int_{-\infty}^{x} \mathrm{d}x \pi R^2 \delta n_c$ profiles are plotted in Fig. S3 for zero and finite Péclet number. These profiles exhibit a strong sensitivity to applied pressures when $\mathrm{Pe}_P < 50$, with the concentration everywhere relaxing to the reservoir value as the linearizing Péclet number $\mathrm{Pe}_P^{\mathrm{lin}} \approx 300$ is approached. The structural similarity between the cumulative charge and the excess concentration relative to the reservoir value $\delta c \equiv c - 1$ is immediately apparent and illustrates the direct relationship between the concentration field and the net spatial charge $\delta n_c$.

### Simplification of Axisymmetric Poisson–Nernst–Planck–Stokes Equations in a Conical Geometry

The full axisymmetric steady-state Poisson–Nernst–Planck–Stokes (PNPS) equations are

$$j_\pm^x = D\left[(-\partial_x c_\pm) \pm \frac{e}{k_B T} c_\pm (-\partial_x \phi)\right] + u c_\pm, \qquad \textbf{[S1]}$$

$$j_\pm^r = D\left[(-\partial_r c_\pm) \pm \frac{e}{k_B T} c_\pm (-\partial_r \phi)\right] + w c_\pm, \qquad \textbf{[S2]}$$

$$\partial_x j_\pm^x + \frac{\partial_r (r j_\pm^r)}{r} = 0, \qquad \textbf{[S3]}$$

$$\epsilon_r \epsilon_0 \left[\partial_x^2 \phi + \frac{\partial_r (r \partial_r \phi)}{r}\right] + n_c = 0, \qquad \textbf{[S4]}$$

$$\partial_x P + n_c \partial_x \phi = \eta \left[\partial_x^2 u + \frac{\partial_r (r \partial_r u)}{r}\right], \qquad \textbf{[S5]}$$

$$\partial_r P + n_c \partial_r \phi = \eta \left[\partial_x^2 w + \frac{\partial_r (r \partial_r w)}{r} - \frac{w}{r^2}\right], \quad \text{and} \qquad \textbf{[S6]}$$

$$\partial_x u + \frac{\partial_r (r w)}{r} = 0. \qquad \textbf{[S7]}$$

In the above, $j_\pm^x$ and $j_\pm^r$ are, respectively, the longitudinal and radial ionic fluxes, and $u$ and $w$ are the longitudinal and radial hydrodynamic velocities. All other variables and coefficients are defined in the main text.

We partition the ionic charge $n_c \equiv e(c_+ - c_-)$ into a component satisfying a local Poisson-Boltzmann (PB) equilibrium, $n_c^{\mathrm{PB}}(x, r)$, and a deviation from local equilibrium, $\delta n_c(x)$, which a priori need not be confined to the Debye layer and is thus assumed to be radially uniform. PB theory requires local electroneutrality when summing over the contributions of the ions and surface to the total charge, so that the local equilibrium charge density must satisfy $\langle n_c^{\mathrm{PB}} \rangle + 2\sigma/R = 0$, where the angled brackets indicate a cross-sectional average, $\sigma$ is the surface charge density at the solid–liquid interface, and $R$ is the local radius. Thus, $\delta n_c(x) \equiv \langle n_c \rangle + 2\sigma/R$ represents the density of net spatial charge at a location $x$ along the nanopore. We assume that the net spatial charge is small compared with the surface charge, i.e., that $|\delta n_c| \ll |\langle n_c^{\mathrm{PB}} \rangle| = 2|\sigma|/R$. In particular, this implies that, within the Debye layer where the equilibrium ionic charge is significant, $n_c \approx n_c^{\mathrm{PB}}$, whereas outside of the Debye layer (in the bulk), $n_c \approx \delta n_c$.

Eqs. **S1**–**S7** are dramatically simplified if we assume that the longitudinal scale of variation of $c_+$, $c_-$, $\phi$, and $P$ is large compared with the minimum radius $R_0$ (1, 2). This assumption is confirmed by our theoretical results (Fig. S3). In this case, we may neglect longitudinal gradients, except in the forcing terms $\partial_x \phi$ and $\partial_x P$, and the radial velocity. We find

$$j_\pm^x = D\left[(-\partial_x c_\pm) \pm \frac{e}{k_B T} c_\pm (-\partial_x \phi)\right] + u c_\pm, \qquad \textbf{[S8]}$$

$$\epsilon_r \epsilon_0 \frac{\partial_r (r \partial_r \delta\phi)}{r} + n_c^{\mathrm{PB}} = 0, \qquad \textbf{[S9]}$$

$$\delta c = c_0 \left[\cosh\left(\frac{e\delta\phi}{k_B T}\right) - 1\right], \qquad \textbf{[S10]}$$

$$n_c^{\mathrm{PB}} = -e c_0 \sinh\left(\frac{e\delta\phi}{k_B T}\right), \qquad \textbf{[S11]}$$

$$\partial_x P + n_c d_x \phi_0 + n_c^{\mathrm{PB}} \partial_x \delta\phi = \eta \frac{\partial_r (r \partial_r u)}{r}, \quad \text{and} \qquad \textbf{[S12]}$$

$$\partial_r P + n_c^{\mathrm{PB}} \partial_r \delta\phi = 0. \qquad \textbf{[S13]}$$

In the above, we have taken advantage of the fact that $\lambda_D / R_0 \ll 1$ to partition the concentration and electrostatic potential as $c(x, r) = c_0(x) + \delta c(x, r)$ and $\phi(x, r) = \phi_0(x) + \delta\phi(x, r)$, where $c_0(x) \equiv c(x = 0, r)$ and $\phi_0(x) \equiv \phi(x = 0, r)$ are the ionic concentrations and electrostatic potential along the centerline of the nanopore (1, 2). We refer to the centerline concentration $c_0(x)$ as the concentration $c$ in the main text and in *Supporting Information*, except for this section and the following one. $\delta c(x, r)$ and $\delta\phi(x, r)$ then represent the radial deviations in concentration and potential induced by the Debye layer. In Eqs. **S12** and **S13** we have also made use of the fact that $\delta\phi$ is substantially different from zero only within the Debye layer to make the approximation $n_c \nabla \delta\phi \approx n_c^{\mathrm{PB}} \nabla \delta\phi$.

Eqs. **S9**–**S11** describe the local PB quasi-equilibrium holding in the radial direction at each point $x$ along the nanopore. Combining the Boltzmann distribution obtained in Eqs. **S10** and **S11** with the radial momentum balance, Eq. **S13**, and integrating gives an osmotic pressure balance in the radial direction (3),

$$P(x, r) = P_0(x) + k_B T \delta c(x, r), \qquad \textbf{[S14]}$$

where $P_0(x) \equiv P(x = 0, r)$ is the pressure at the nanopore centerline. Inserting Eq. **S14** into Eq. **S12** gives

$$(d_x P_0 + \delta n_c d_x \phi_0) + \left(n_c^{\mathrm{PB}} d_x \phi_0\right) + \left(k_B T \frac{\delta c}{c_0} d_x c_0\right) = \eta \frac{\partial_r (r \partial_r u)}{r}, \qquad \textbf{[S15]}$$

where, on the left-hand side, we have segregated the forcing terms into, from left to right, the pressure gradient and electric body force associated with the net spatial charge, which together induce a quadratic Hagen–Poiseuille flow, and the electroosmotic and diffusioosmotic driving forces, which both induce plug-like flows.

## Derivation of Eqs. 4–8: Radial Integration of PNPS Equations

In this section, we integrate the simplified axisymmetric PNPS equations (Eqs. **S8**–**S11** and Eq. **S15**) in the radial direction to obtain Eqs. **4**–**8** in the main text. We retain dimensioned variables until the final result, where we nondimensionalize as in the main text.

Partitioning the advective terms into bulk and surface components, we may write the ionic current $I \equiv \int \mathrm{dA}\, e\, (j_+^x - j_-^x)$ and solute flux $J_{\mathrm{sol}} \equiv \int \mathrm{dA}\, (j_+^x + j_-^x)$ as

$$I = \pi R^2 D \left[ (-\partial_x \langle n_c \rangle) + \frac{e^2}{k_B T} (-\langle c \partial_x \phi \rangle) \right] + Q \delta n_c + \int \mathrm{dA}\, u n_c^{\mathrm{PB}}, \qquad \textbf{[S16]}$$

and

$$J_{\mathrm{sol}} = \pi R^2 D \left[ (-\partial_x \langle c \rangle) + (-\langle n_c \partial_x \phi \rangle) \right] + Q c_0 + \int \mathrm{dA}\, u \delta c, \qquad \textbf{[S17]}$$

where in the diffusive terms we have taken advantage of the fact that the geometry is slowly varying everywhere except in the vicinity of the tip to transpose the order of the derivative and cross-sectional average. In the above, $Q \equiv \int \mathrm{dA}\, u$ is the total solvent flux, the sum of contributions from the Hagen–Poiseuille, electroosmotic, and diffusioosmotic flows.

We simplify the electrodiffusive terms in Eqs. **S16** and **S17** by making two further simplifying approximations: (*i*) that the average of the nonlinear electrophoretic terms may be approximated as the product of the averages, i.e., $\langle c_\pm \partial_x \phi \rangle \approx \langle c_\pm \rangle \langle \partial_x \phi \rangle \approx \langle c_\pm \rangle \partial_x \langle \phi \rangle$, and (*ii*) that the averages of the total ionic concentration and electrostatic potential are approximately equal to their centerline values, $\langle c \rangle \approx c_0$ and $\langle \phi \rangle \approx \phi_0$. The former assumption is supported by the qualitative and quantitative success of one-dimensional Poisson–Nernst–Planck models in reproducing observed IV behavior in nanofluidic devices (4–6), and the latter is justified by the fact that the Dukhin number is of order unity, meaning the additional concentration and electrostatic potential in the Debye layer induced by the surface charge are roughly of the order of the values in the bulk (7), but they are confined to a region of area $\lambda_D \times 2\pi R \ll \pi R^2$ and thus do not contribute significantly to the cross-sectional average.

From the radial integration of the longitudinal Stokes equation (Eq. **S15**) with the local PB equilibrium equations, Eqs. **S9**–**S11**, one may derive expressions for the three contributions to the total solvent flux, the Hagen–Poiseuille (8), electroosmotic (7, 8), and diffusioosmotic (3, 9) solvent fluxes $Q_{\mathrm{HP}}$, $Q_{\mathrm{EO}}$, and $Q_{\mathrm{DO}}$, respectively. These calculations are conceptually straightforward but lengthy and are performed in Refs. 3, 7, 8, and 9. We thus quote only the results here:

$$Q_{\mathrm{HP}} = \frac{\pi R^4}{8\eta} (-d_x P_0 - \delta n_c d_x \phi_0), \qquad \textbf{[S18]}$$

$$Q_{\mathrm{EO}} = \mu_{\mathrm{EO}} \pi R^2 (-d_x \phi_0), \quad \text{and} \qquad \textbf{[S19]}$$

$$Q_{\mathrm{DO}} = \mu_{\mathrm{DO}} \pi R^2 (-d_x \ln c_0). \qquad \textbf{[S20]}$$

In Eqs. **S19** and **S20**, we have retained only the terms that are lowest order in the parameter $(\lambda_D / R_0)/\mathrm{Du}_0 \equiv \ell_{\mathrm{GC}} / \lambda_D$, where $\ell_{\mathrm{GC}} \equiv 2\lambda_D^2 e c_0 / |\sigma|$ is the Gouy–Chapman length. This is consistent with the experiments, which are conducted in the regime that the Debye length is small compared with the tip radius, but the Dukhin number is of order unity.

Finally, one may calculate the advective surface current $\int \mathrm{dA}\, u n_c^{\mathrm{PB}}$ by using the Hagen–Poiseuille, diffusioosmotic, and electroosmotic velocity profiles obtained from radial integration of Eq. **S15** along with the PB charge distribution $n_c^{\mathrm{PB}}$ given in Eq. **S11**. One thus obtains expressions for the streaming current (7, 8), associated with the pressure-driven flow in the Debye layer—and the electroosmotic (7, 9) and diffusioosmotic (9) currents—denoted, respectively, $I_{\mathrm{stm}}$, $I_{\mathrm{EO}}$, and $I_{\mathrm{DO}}$. We again quote only the results here:

$$I_{\mathrm{stm}} = \mu_{\mathrm{EO}} \pi R^2 (-d_x P_0), \qquad \textbf{[S21]}$$

$$I_{\mathrm{EO}} = 4\pi R |\sigma| \frac{\epsilon_r \epsilon_0}{\eta} \frac{k_B T}{e} (-d_x \phi_0), \quad \text{and} \qquad \textbf{[S22]}$$

$$I_{\mathrm{DO}} = 4\pi R \sigma \frac{\epsilon_r \epsilon_0}{\eta} \left( \frac{k_B T}{e} \right)^2 d_x \ln c_0. \qquad \textbf{[S23]}$$

In Eqs. **S22** and **S23** we have again retained only terms of leading order in $(\lambda_D / R_0)/\mathrm{Du}_0$.

We note that similar surface transport terms exist for the concentration; however, as the bulk concentration $c_0$ is of order unity and the Debye length is small compared with the nanopore radius, these terms will be dwarfed by the bulk solute advection and are thus neglected.

From Eqs. **S16–S23** and the approximations noted above, we find

$$
\begin{aligned}
I &= \pi R^2 D\left[(-\partial_x\langle n_c\rangle) + \frac{e^2}{k_B T}(-c_0 d_x\phi_0)\right] \\
&\quad + Q\delta n_c + \mu_{\mathrm{EO}}\pi R^2(-d_x P_0) \\
&\quad + 4\pi R|\sigma|\frac{\epsilon_r\epsilon_0}{\eta}\frac{k_B T}{e}\left(-d_x\phi_0 + \mathrm{S}\frac{k_B T}{e}d_x \ln c_0\right), \\
J_{\mathrm{sol}} &= \pi R^2 D\left[(-\partial_x c_0) + (-\langle n_c\rangle\partial_x\phi_0)\right] + Qc_0, \quad \text{and} \\
Q &= \frac{\pi R^4}{8\eta}(-d_x P_0 - \delta n_c d_x\phi_0) + \mu_{\mathrm{EO}}\pi R^2(-d_x\phi_0) \\
&\quad + \mu_{\mathrm{DO}}\pi R^2(-d_x \ln c_0).
\end{aligned}
$$

Dropping the subscript zeros and the angled brackets on $n_c$, and nondimensionalizing as in the main text, we obtain

$$
I = \pi R^2\left(-\frac{dn_c}{dx}\right) + \pi R^2 c\left(-\frac{d\phi}{dx}\right) + Q\delta n_c + \frac{1}{\mathrm{Pe}_{\mathrm{osm}}}\frac{\mu_{\mathrm{EO}}}{\mu_{\mathrm{EP}}} \times\pi R^2\left(-\frac{dP}{dx}\right) + 2\pi Rk\left(-\frac{d\phi}{dx}\right) + 2\pi Rk\left(\mathrm{S}\frac{d\ln c}{dx}\right), \quad \textbf{[S24]}
$$

$$
J_{\mathrm{sol}} = -\pi R^2\left(\frac{dc}{dx} + n_c\frac{d\phi}{dx}\right) + Qc, \quad \textbf{[S25]}
$$

$$
Q = -\pi R^4\left(\frac{dP}{dx} + \mathrm{Pe}_{\mathrm{osm}}\delta n_c\frac{d\phi}{dx}\right) + \pi R^2\left[\frac{\mu_{\mathrm{EO}}}{\mu_{\mathrm{EP}}}\left(-\frac{d\phi}{dx}\right) + \frac{\mu_{\mathrm{DO}}}{D}\left(-\frac{d\ln c}{dx}\right)\right], \quad \textbf{[S26]}
$$

i.e., Eqs. **4–6** in the main text.

Finally, the appropriate conservation equations for the ionic current and the solute, solvent, and electric fluxes are obtained by integrating Eqs. **S3**, **S4**, and **S7** over a longitudinally infinitesimal control volume, again making use of the slowly varying geometry, and making the approximation $\langle\phi\rangle \approx \phi_0$. The procedure is standard, and we do not recapitulate it here. The result, in nondimensionalized variables, is

$$
\frac{dI}{dx} = \frac{dJ}{dx} = \frac{dQ}{dx} = 0, \quad \text{and} \quad \textbf{[S27]}
$$

$$
\left(\frac{\lambda_D}{\ell}\right)^2\frac{1}{\pi R^2}\frac{d}{dx}\left(\pi R^2\frac{d\phi}{dx}\right) + \delta n_c = 0. \quad \textbf{[S28]}
$$

## Derivation of Eqs. 11 and 12

We first derive Eqs. **11** and **12** from the main text, relating the structure of the SCZ to the ionic current.

We begin by integrating the Poisson equation, Eq. **S28**, to determine an expression for the electrostatic potential in terms of $\delta n_c$. We find that the boundary conditions on $\phi$ require the limit of the electric flux far in the interior of the nanopore to be finite, $\lim_{x\to-\infty}\pi R^2(-d_x\phi) \neq 0$. As $c(x \to -\infty) = 1$, we recognize this limiting electric flux as the electrophoretic current, $I_{\mathrm{ep}} = \pi R^2 c(-d_x\phi)$, in the interior of the nanopore–i.e., $\lim_{x\to-\infty}\pi R^2(-d_x\phi) = I_{\mathrm{ep}}(-\infty)$. Thus, we may integrate Eq. **S28** twice and solve for $I_{\mathrm{ep}}(-\infty)$ to obtain

$$
I_{\mathrm{ep}}(-\infty) = G_b\left[\Phi - \left(\frac{\lambda_D}{\ell}\right)^{-2}\int_{-\infty}^{+\infty}\frac{\mathrm{dx}}{\pi R^2}\int_{-\infty}^{x}\mathrm{dx}'\pi R^2\delta n_c\right], \quad \textbf{[S29]}
$$

where $G_b$ is the bulk electrophoretic conductance, as defined in the main text.

Further, the constancy of the limiting electric flux implies, from Eq. **S28**, that $\delta n_c$ must vanish faster than $1/R^2$ as $x \to -\infty$. Thus, $n_c \to -S(2\mathrm{Du}_0/R)$ and the diffusive current in the interior is given by

$$
I_{\mathrm{diff}}(-\infty) = \mathrm{S}2\pi\frac{\ell}{R_0}\alpha_1\mathrm{Du}_0, \quad \textbf{[S30]}
$$

where we have used the fact that $dR/dx \to -\frac{\ell}{R_0}\alpha_1$.

Examining the limiting behavior as $x \to -\infty$ of $J_{\mathrm{sol}}$ and $Q$ (Eqs. **S25** and **S26**, respectively), we find that $dc/dx$ must vanish as $1/R^2$ or faster and $dP/dx$ must vanish as $1/R^4$ or faster to ensure finite solute and solvent fluxes. We therefore find from the equation for the ionic current (Eq. **S24**) evaluated at $x \to -\infty$ that only the diffusive and electrophoretic contributions are nonvanishing there. Therefore, the total current must equal the sum of the diffusive and electrophoretic currents far in the nanopore interior, irrespective of the value of $\Phi$ or $\mathrm{Pe}_P$. That is, $I = I(-\infty) = I_{\mathrm{diff}}(-\infty) + I_{\mathrm{ep}}(-\infty)$. From this, we see that the diffusive term proportional to $\mathrm{Du}_0\alpha_1$ must be compensated for by the formation of an SCZ when $\Phi = 0$ to ensure zero flux at equilibrium. This requires that the equilibrium net charge density, $\delta n_c^{\mathrm{equ}}$, be related to the surface charge and geometry by

$$
G_b\left(\frac{\ell}{\lambda_D}\right)^2\int_{-\infty}^{+\infty}\frac{\mathrm{dx}}{\pi R^2}\int_{-\infty}^{x}\mathrm{dx}'\pi R^2\delta n_c^{\mathrm{equ}} = 2\pi\frac{\ell}{R_0}\alpha_1\mathrm{S}\,\mathrm{Du}_0, \quad \textbf{[S31]}
$$

Eq. **11** from the main text. Combining Eqs. **S29–S31**, we obtain for the ionic current

$$
I = G_b\left[\Phi - \left(\frac{\ell}{\lambda_D}\right)^2\int_{-\infty}^{+\infty}\frac{\mathrm{dx}}{\pi R^2}\int_{-\infty}^{x}\mathrm{dx}'\pi R^2\left[\delta n_c - \delta n_c^{\mathrm{equ}}\right]\right], \quad \textbf{[S32]}
$$

which is Eq. **12** from the main text.

## Derivation of Eq. 13

The results of the full numerical calculations indicate that, when $\Phi \neq 0$ and $\mathrm{Pe}_P = 0$, the current in the nanopore interior is composed primarily of the local electrophoretic and -osmotic contributions. Retaining only these terms, we obtain $I \approx \pi R^2(c + 2k/R)(-d_x\phi)$, and we integrate to find for the apparent conductance $G \equiv I/\Phi$

$$
G = \left[\int_{-\infty}^{+\infty}\frac{\mathrm{dx}}{\pi R^2\left(c + \frac{2k}{R}\right)}\right]^{-1}, \quad \textbf{[S33]}
$$

recovering Eq. **13** of the main text.

## Derivation of Eq. 14

At high pressure, the solvent flux is dominated by the pressure-driven (Hagen–Poiseuille) component: $Q \approx \pi R^4(-d_x P)$ (Eq. **S26**). We integrate this to find $Q = \left(\int_{-\infty}^{+\infty}\mathrm{dx}/\pi R^4\right)^{-1}\mathrm{Pe}_P$. The pressure gradient in the nanopore is thus approximately given by

$$
-\frac{dP}{dx} = \frac{Q}{\pi R^4} = \left(\int_{-\infty}^{+\infty}\frac{\mathrm{dx}}{\pi R^4}\right)^{-1}\frac{\mathrm{Pe}_P}{\pi R^4}. \quad \textbf{[S34]}
$$

The streaming current $I_{\mathrm{stm}} = \mathrm{Pe}_{\mathrm{osm}}^{-1}(\mu_{\mathrm{EO}}/\mu_{\mathrm{EP}})\pi R^2(-d_x P)$ (Eq. **S24**) is thus given by

$$
I_{\mathrm{stm}} = \left(\int_{-\infty}^{+\infty}\frac{\mathrm{dx}}{\pi R^4}\right)^{-1}\frac{\mu_{\mathrm{EO}}}{\mu_{\mathrm{EP}}}\frac{\mathrm{Pe}_P}{\mathrm{Pe}_{\mathrm{osm}}}\frac{1}{R^2}. \quad \textbf{[S35]}
$$

At high $\mathrm{Pe}_P$, the concentration is homogenized—i.e., $c \equiv 1$ (Fig. S3). The electrophoretic current is therefore equal to the electric flux, $I_{\mathrm{ep}}(x) = \pi R^2(-d_x\phi)$, and we may substitute this into the Poisson equation, Eq. **S28**, and integrate once to obtain

$$I_{\mathrm{ep}}(x) = I + \left(\frac{\ell}{\lambda_D}\right)^2 \int_{-\infty}^{x} \mathrm{d}x' \pi R^2 \delta n_c, \qquad \textbf{[S36]}$$

where we have made the approximation that $I \approx I_{\mathrm{ep}}(-\infty)$ because at high forcing the diffusive current far in the nanopore interior is negligibly small. On the other hand, the numerical results indicate that at high $\mathrm{Pe}_P$ the total current in the nanopore is given principally by the sum of the local electrophoretic, electroosmotic, and streaming currents, $I = I_{\mathrm{ep}} + I_{\mathrm{eo}} + I_{\mathrm{stm}} = (1 + 2k/R)I_{\mathrm{ep}} + I_{\mathrm{stm}}$, where we have again used the fact that $c \equiv 1$. We solve this relation for $I_{\mathrm{ep}}$, insert it into Eq. **S36**, divide the result by $\pi R^2$, and integrate in $x$ to obtain

$$\left(\frac{\ell}{\lambda_D}\right)^2 \int_{-\infty}^{+\infty} \frac{\mathrm{d}x}{\pi R^2} \int_{-\infty}^{x} \mathrm{d}x' \pi R^2 \delta n_c = \left(\frac{1}{G_0} - \frac{1}{G_b}\right) I - \gamma \frac{\mu_{\mathrm{EO}}}{\mu_{\mathrm{EP}}} \frac{\mathrm{Pe}_P}{\mathrm{Pe}_{\mathrm{osm}}}, \qquad \textbf{[S37]}$$

where $G_0$ and $\gamma$ have been introduced in the main text; we recall that they are constants for a given geometry and electroosmotic conductivity $k$.

Finally, this expression may be inserted into Eq. **S32** to obtain Eq. **14** of the main text:

$$I = G_0 \left(\Phi + \gamma \frac{\mu_{\mathrm{EO}}}{\mu_{\mathrm{EP}}} \frac{\mathrm{Pe}_P}{\mathrm{Pe}_{\mathrm{osm}}}\right) \equiv G_0 \Phi + \mathrm{S}_{\mathrm{stm}} \mathrm{Pe}_P. \qquad \textbf{[S38]}$$

## Model Geometry

We impose a continuous exponential variation between the two limiting slopes, $-\alpha_1$ and $+\alpha_2$ in the model of the form (in dimensioned variables)

$$\frac{R(x)}{R_0} = \frac{1 - \alpha_1 \left[(x/\ell) + k_1\right]}{1 - e^{2[(x/\ell)+k_1]}} + \frac{1 + \alpha_2 \left[(x/\ell) + k_1\right]}{1 - e^{-2[(x/\ell)+k_1]}} + k_2, \qquad \textbf{[S39]}$$

where $k_1$ and $k_2$ are constants chosen such that $R(0)/R_0 = 1$. This relation defines the geometric length-scale $\ell$, which is an e-folding scale associated with the transition between limiting slopes.

## Numerical Scheme

To obtain a numerical solution to the steady-state equations $dI/dx = dJ/dx = dQ/dx = 0$ with the ionic current and solute and solvent fluxes given in Eqs. **4–6** in the main text and Eqs. **S24–S26** above, we have used a simple temporally explicit (forward difference) integration scheme, with the spatial gradients approximated by a second-order centered-difference scheme, to evolve the total ionic concentration and net spatial charge according to the physical dynamics. The temporal dependence is given by

$$\partial_t c + \frac{\partial_x J}{\pi R^2} = 0, \quad \text{and} \qquad \textbf{[S40]}$$

$$\partial_t \delta n_c + \frac{\partial_x I}{\pi R^2} = 0. \qquad \textbf{[S41]}$$

Initially, and at each time step, the electrostatic potential and the pressure field and solvent flux are found by numerically integrating, respectively, the Poisson equation and steady-state continuity equation $dQ/dx = 0$ with $Q$ as given above and in the main text. Formal integration of these equations gives

$$Q = \left(\int_{-\infty}^{+\infty} \frac{\mathrm{d}x}{\pi R^4}\right)^{-1} \left\{ \left(\mathrm{Pe}_P + \mathrm{Pe}_{\mathrm{osm}} \int_{-\infty}^{\infty} \mathrm{d}x \delta n_c E\right) + \int_{-\infty}^{+\infty} \mathrm{d}x \left[\frac{\mu_{\mathrm{EO}}}{\mu_{\mathrm{EP}}} E + \frac{\mu_{\mathrm{DO}}}{D} (-\partial_x \ln c)\right] \right\}, \qquad \textbf{[S42]}$$

$$P(x) = \mathrm{Pe}_P - Q \int_{-\infty}^{x} \frac{\mathrm{d}x'}{\pi R^4} + \mathrm{Pe}_{\mathrm{osm}} \int_{-\infty}^{x} \mathrm{d}x' \delta n_c E + \int_{-\infty}^{x} \mathrm{d}x' \left[\frac{\mu_{\mathrm{EO}}}{\mu_{\mathrm{EP}}} E + \frac{\mu_{\mathrm{DO}}}{D} (-\partial_x \ln c)\right], \qquad \textbf{[S43]}$$

$$\mathcal{F}_D(-\infty) = \left(\int_{-\infty}^{+\infty} \frac{\mathrm{d}x}{\pi R^2}\right)^{-1} \left[\left(\frac{\lambda_D}{\ell}\right)^2 \Phi - \int_{-\infty}^{+\infty} \frac{\mathrm{d}x}{\pi R^2} \int_{\infty}^{x} dx' \pi R^2 \delta n_c\right], \quad \text{and} \qquad \textbf{[S44]}$$

$$\phi(x) = \Phi - \left(\frac{\ell}{\lambda_D}\right)^2 \left[\mathcal{F}_D(-\infty) \int_{-\infty}^{x} \frac{\mathrm{d}x'}{\pi R^2} + \int_{-\infty}^{x} \frac{\mathrm{d}x'}{\pi R^2} \int_{\infty}^{x} dx'' \pi R^2 \delta n_c\right], \qquad \textbf{[S45]}$$

where $\mathcal{F}_D(-\infty) \equiv (\lambda_D/\ell)^2 I_{\mathrm{ep}}(-\infty) \equiv (\lambda_D/\ell)^2 \lim_{x\to-\infty} \pi R^2 E$ is the electric displacement flux, and $E \equiv -\partial_x \phi$ is the electric field.

We evolve the concentration and net spatial charge profiles from initial conditions given by

$$c(x, t=0) = \sqrt{\left(\frac{2\mathrm{Du}_0}{R}\right)^2 + 1}, \quad \text{and} \qquad \textbf{[S46]}$$

$$n_c(x, t=0) = \frac{2\mathrm{Du}_0}{R}, \qquad \textbf{[S47]}$$

with the initial values of $Q$, $P$, and $\phi$ calculated according to Eqs. **S42–S45** above. Eqs. **S46** and **S47** come from imposition of electroneutrality and the PB equilibrium condition $c_+ c_- = 1$. Numerical integration continues until the solute flux and ionic concentration relax to steady, spatially uniform values.

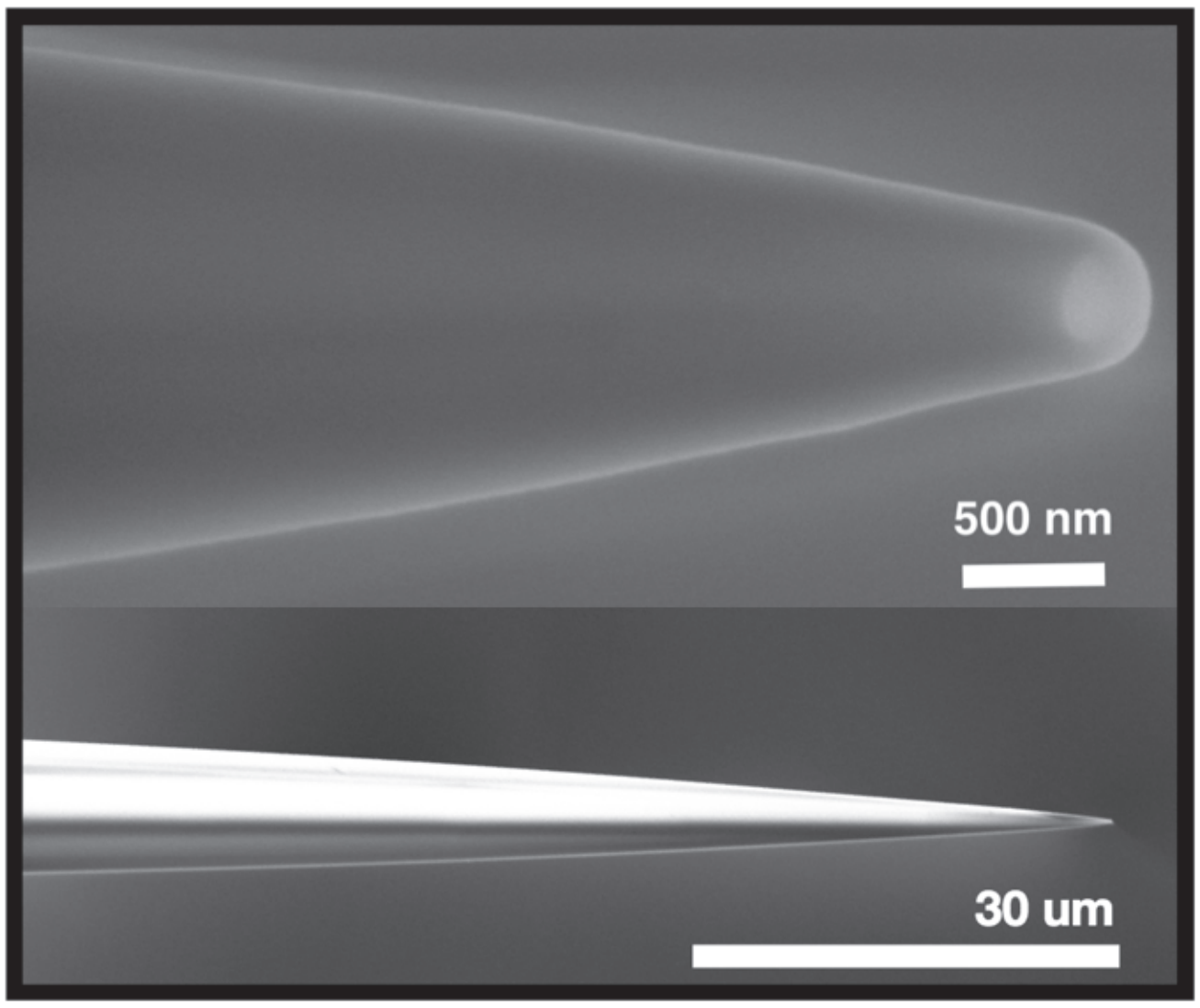

**Fig. S1.** SEM images of the tip of the glass nanopipette. The nanopipette is tilted in *Upper* image to reveal the internal radius at the tip.

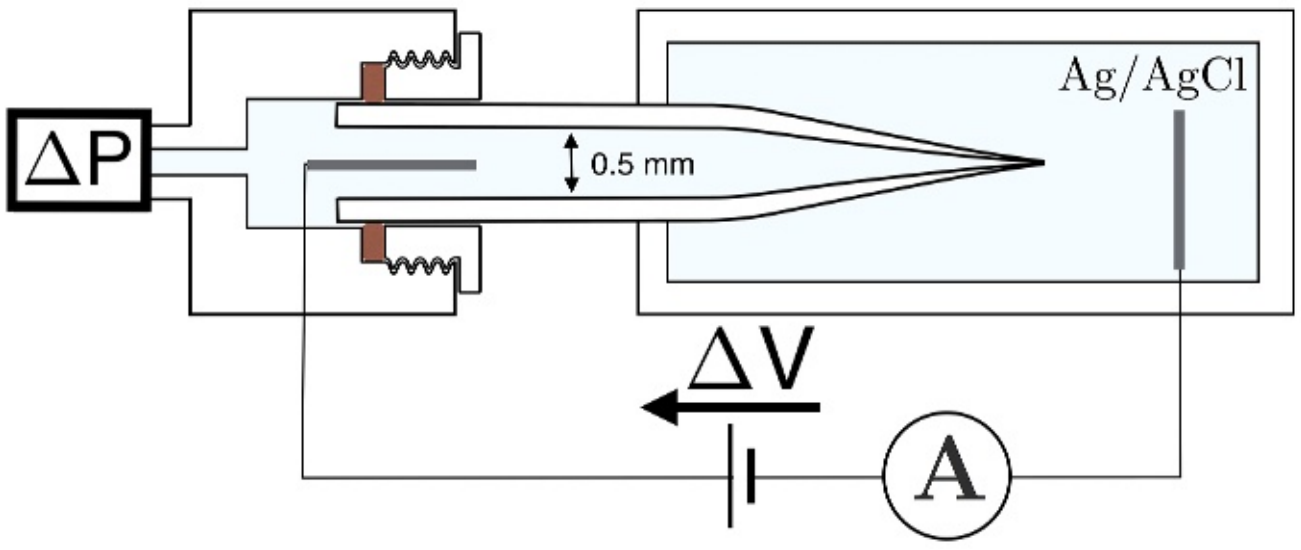

**Fig. S2.** Experimental setup used to measure ionic current in the nanopipette as a function of applied voltage $\Delta V$ and pressure $\Delta P$. Pressure is applied in the reservoir with the working electrode at the millimetric end of the nanopipette (the upstream reservoir). The concentrations and pH in both reservoirs are equal.

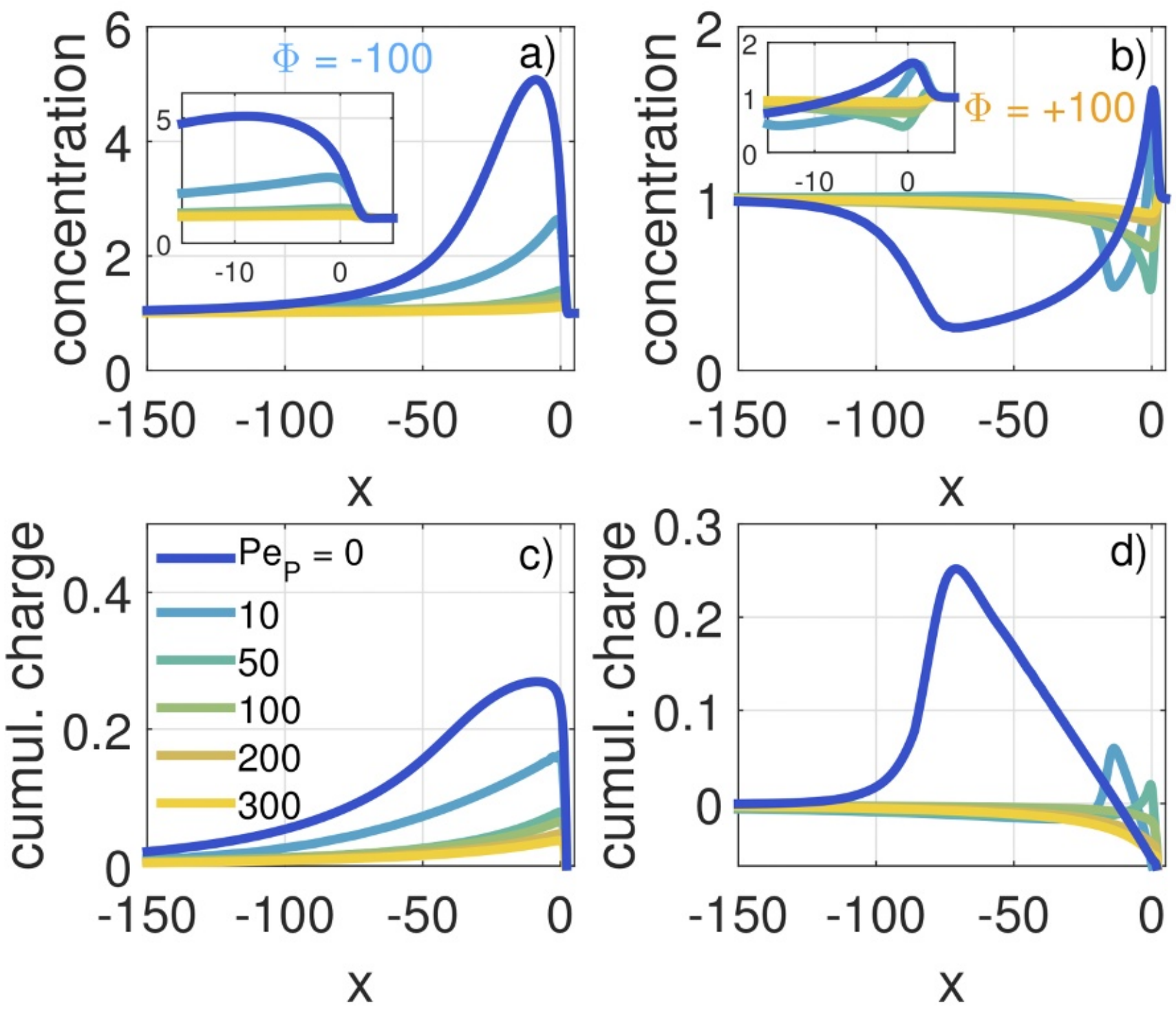

**Fig. S3.** Model-derived profiles of local concentration (*A* and *B*) and cumulative charge (*C* and *D*) for $\Phi = -100$ (*A* and *C*) and $\Phi = +100$ (*B* and *D*). The profiles are colored according to increasing values of $Pe_P$, as indicated in the key in *C*. *Insets* in *A* and *B* show the variations in concentration in the vicinity of $x = 0$.